\documentstyle[pre,preprint,aps]{revtex}
\begin{document}
\title{Errors in Monte Carlo simulations using shift
register random number generators }

\author {F. Schmid and N.B. Wilding}
\address{ Institute f\"{u}r Physik,
Johannes Gutenberg Universit\"at Mainz, D-55099 Mainz, Germany.}

\date{November 1995}
\maketitle
\begin{abstract}

We report large systematic errors in Monte Carlo simulations of the
tricritical Blume-Capel model using single spin Metropolis updating. The
error, manifest as a $20\%$ asymmetry in the magnetisation distribution,
is traced to the interplay between strong triplet correlations in the
shift register random number generator and the large tricritical
clusters.  The effect of these correlations is visible only when the
system volume is a multiple of the random number generator lag
parameter. No such effects are observed in related models.

\end{abstract}
\vfill

\section{Introduction}

Interest in random number generator (RNG) deficiencies has recently been
aroused by the work of Ferrenberg et al \cite{ferrenberg}, who found
that the widely employed ``R250'' generator produces systematic errors
in Monte Carlo simulations using cluster updating algorithms. Their high
precision measurements of the specific heat and the internal energy of
the critical two dimensional Ising model differed from the exact values
by up to $0.05\%$ \cite{ferrenberg}.

These observations prompted a number of detailed studies of systematic
errors in cluster algorithms and other ``depth first'' algorithms, such
as those employed for simulating self avoiding walks
\cite{grassberger}-\cite{vattulainen2}.  The consensus emerging from
these studies was that pseudo random numbers (PRNs) $\{X_n\}$ generated
by generalized feedback shift register (GFSR) algorithms of the type
\cite{RNG}

\begin{equation}
X_n = X_{n-p} \oplus X_{n-q} \qquad (\oplus \equiv \mbox{exclusive-or}),
\label{eq:rng}
\end{equation}
are generally ``bad'' when used in depth first algorithms, even though
they pass many of the standard RNG tests and have very long periods
\cite{vattulainen1}.  However for local (single spin flip) Metropolis
updating, good results were obtained using the GFSR generators. It was
therefore concluded that this type of algorithm is relatively
insensitive to correlations and thus safe to use.

In this paper, we show that in certain circumstances large systematic
errors can also arise when using GFSRs generators in conjunction with
local Metropolis updating. We demonstrate this for the case of the
tricritical Blume-Capel model and argue that the observed errors
originate from certain extremely large triplet correlations in the
random number sequences.

\section{Results}

\begin{figure}[h]
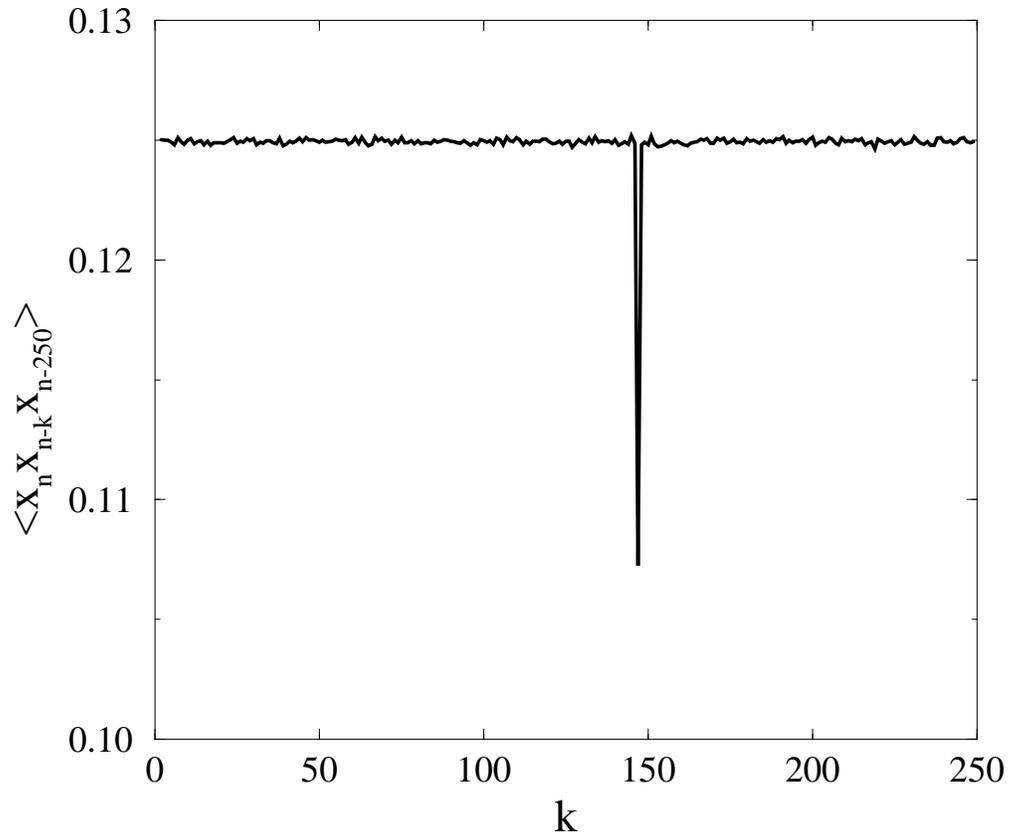

\caption{The triplets $\langle X_nX_{n-k}X_{n-250}\rangle$ as a
function of $1\le k \le 249$, generated from a sequence of $10^8$ PRNs
using the generator of equation~\protect\ref{eq:rng} with $p=250$, $q=147$.}

\label{fig:triplets}
\end{figure}

There have been a number of recent suggestions in the literature that
the problems observed with the GFSR generators may originate from
triplet correlations \cite{grassberger,vattulainen2}. In view of this
we have made a study of three point correlations in the GFSR
generators and have indeed found an extremely large
effect in the $(X_n,X_{n-p},X_{n-q})$ triplet. For this triplet the
three point average \mbox{$\langle X_n \cdot X_{n-p} \cdot X_{n-q}
\rangle$} takes the value $0.107$ instead of $(1/2)^3 = 0.125$. This
effect is displayed in figure~\ref{fig:triplets} where we plot the
triplets $\langle X_nX_{n-k}X_{n-250}\rangle$ as a function of $k$,
for $1\le k \le 249$. The results were obtained from a sequence of
$10^8$ PRNs using the magic number pair $(250,147)$ corresponding to
the R250 generator.  In fact the same effect, with the same magnitude,
is observable for other magic number pairs $(p,q)$ recommended in
the literature \cite{RNG}, viz $(521, 168)$, $(1279,216)$,
$(1279,418)$, $(3217,67)$, $(3217,576)$, $(4423, 1393)$,
$(4423,2098)$, $(9689,84)$, $(9689,471)$, $(9689,1836)$,
$(9689,2444)$, $(9689,4187)$.

It is thus conceivable that triplet correlations are responsible for the
systematic errors observed in ``depth first'' algorithms, which
principally rely on long uncorrelated sequences of random numbers.
However, as mentioned above, it has hitherto been generally believed
that local Metropolis updating schemes are insensitive to
correlations in the PRN sequences and give results at least better than
linear congruential RNGs \cite{ferrenberg,vattulainen1}. In fact we have
found that in one system, the tricritical Blume-Capel model, triplet
correlations can lead to extremely large ($\sim 20\%$) errors, even when
using a Metropolis single site updating scheme.

The Blume-Capel model is a spin 1 Ising model with the Hamiltonian

\begin{equation}
{\cal H}_{BC} = - \sum_{<ij>} s_i s_j + \Delta \sum_i s_i{}^2,
\end{equation}
where spins $s_i$ take the values $s_i = 0, \pm 1$ and the sum $<ij>$ runs
over all nearest neighbor pairs.  We have simulated this model on an FCC
lattice for system size $V=(5 \times 5 \times 5 \cdot 4)=500$ spins.
Spins were updated serially, each update requiring two random numbers:
one to choose the new spin, and one for the Metropolis step. The random
numbers were generated by the R250 RNG, having the  magic number
pair (p=250, q=147). Figure~\ref{fig:asymm}(a) shows the measured
probability distribution of the magnetisation $M=\sum_is_i$ at the
near-tricritical parameters $k_BT = 3.0$, $\Delta = 5.67$ (circles).
Notwithstanding the symmetry of the Hamiltonian under inversion of
all spins ($\{s_i\} \to \{-s_i\}$), the distribution is clearly
asymmetric with respect to the ``plus'' and ``minus'' phases. We find,
however, that if spins are updated randomly instead of serially, the
distribution becomes symmetric [figure~\ref{fig:asymm}(b)].

We believe that the observed asymmetry in $P(M)$ is related to the
fact that for system size $V=500$, the larger generator lag $p=250$ is a
factor of the system volume, so that a given spin is always updated
with a random number having the same relative position within successive
sequences of $250$ PRNs. This view is supported by the following evidence:

\begin{itemize}

\item We find that the asymmetry in $P(M)$ disappears when other
system sizes are simulated, e.g. $(14\times 14 \times 14 \cdot 4) =
1372$.

\item The asymmetry disappears for the $V=500$ system size when one
additional PRN is drawn and discarded after each Monte
Carlo step (i.e. after a whole sweep through the lattice).

\item The asymmetry disappears for the $V=500$ system size when the
R250 generator is substituted by the ``R1279'', a GFSR RNG with the magic
number pair (1279,216).

\item The asymmetry reappears with the R1279, for the $V=500$
system size when $279$ additional PRNs are drawn and discarded after
each Monte Carlo step.

\end{itemize}

We further believe that the asymmetry in $P(M)$ is a manifestation of
the aforementioned triplet correlations in the PRN sequence
$(X_n,X_{n-p},X_{n-q})$. To substantiate this we performed a simulation
using the lagged Fibonacci RNG \cite{marsaglia}

\begin{equation}
X_n = X_{n-p} + X_{n-q} \quad \mbox{mod} \; 2^{31} \qquad \mbox{with} \quad
(p,q) = (1279,216),
\end{equation}
which is formally similar to the ``R1279'', but does not display the
triplet correlation. Even after discarding $279$ PRNs after each
cycle, $P(M)$ remains symmetric with this generator.

It is interesting to observe that a number of further factors influence
the size of the asymmetry effect. One is the proximity to the
tricritical point. If we choose $\Delta = 0, k_B T = 6.8$, corresponding
to the critical point of the spin 1 Ising model, the effect weakens, but
remains clearly discernible (not shown).  We also note that the
asymmetry effect is not influenced by the dimensionality or the type of
lattice: Simulations of the $L=10\times 10$ Blume-Capel model on a
square lattice (with $50$ additional random numbers discarded between
each sweep) exhibit the asymmetry with a similar magnitude to that
observed in the three-dimensional case.

\begin{figure}[h]
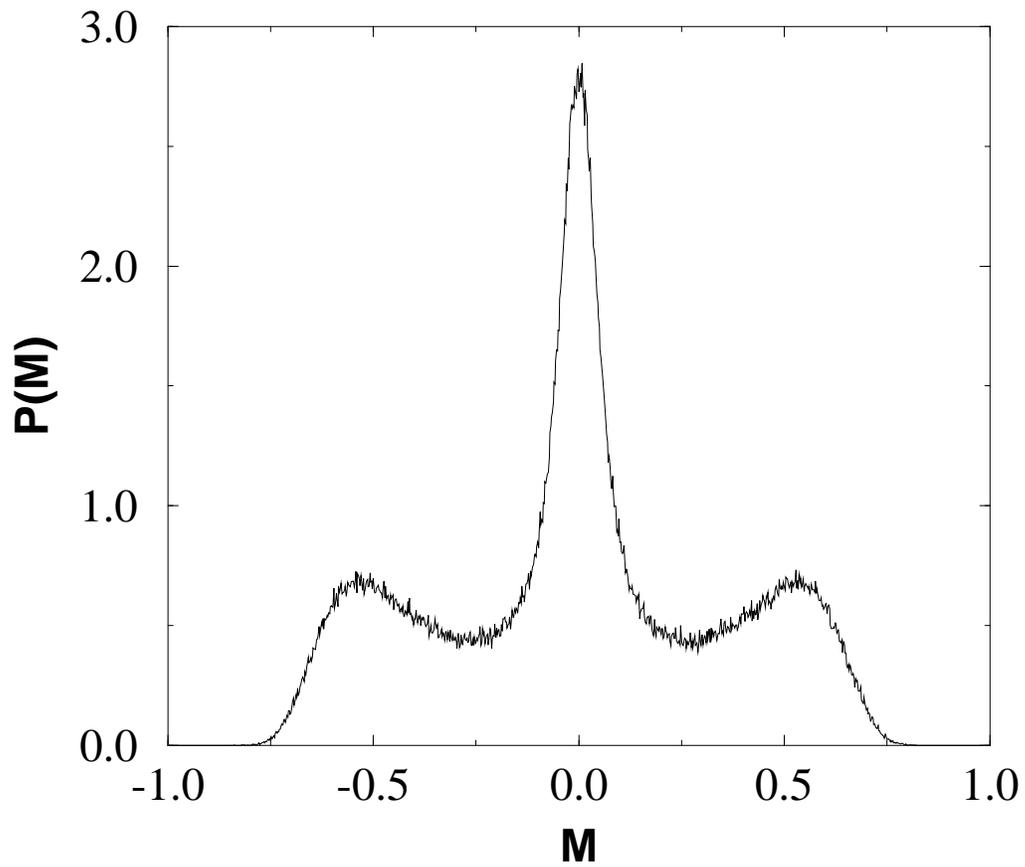


\caption{Magnetisation distribution of the three state Potts model on the
$V=500$ FCC lattice at $k_B T=3.9$, obtained with PRNs generated by
the R250 RNG.}

\label{fig:potts}
\end{figure}

The effect does, however, seem to depend on the symmetry of the
Hamiltonian of the model studied.  To demonstrate this we have also
performed simulations on the three state Potts model near its critical
point. This is a lattice spin model with the Hamiltonian

\begin{equation} {\cal H}_P = - \sum_{<ij>} \delta_{S_i,S_j},
\end{equation}
where $\delta_{ij}$ denotes the Kronecker symbol and $s_i$ takes again
the values $s_i = 0,\pm 1$. The sole alteration required to the
Blume-Capel program to simulate this model, is the calculation of the
energy change in the Metropolis step. In figure~\ref{fig:potts} we
show the magnetisation distribution $P(M)$ for the three state Potts
model on a $V=500$ FCC lattice, for the near-critical temperature
$k_BT=3.9$. No asymmetry between the ``plus'' and ``minus'' phases is
evident. Similar simulations on the 2D simple Ising model also reveal
no asymmetry.

\begin{figure}[h]
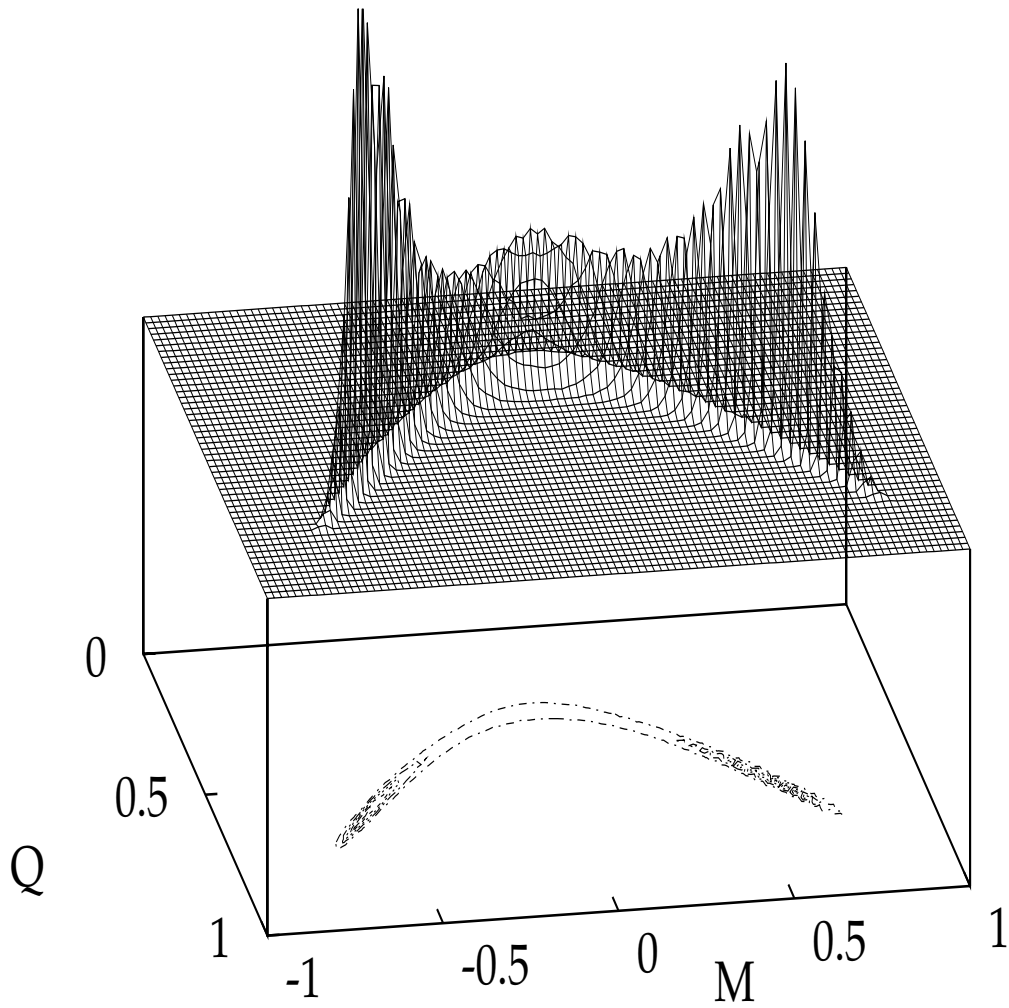


\caption{The joint distribution $P(M,Q)$ of the near-tricritical
Blume-Capel model.  Also shown on the grid base is a contour plot of
the histogram surface.}

\label{fig:contours}
\end{figure}

It is thus clear that the observed asymmetry in the tricritical
Blume-Capel model is the result of a subtle interplay between the
tricritical clusters and PRN triplet correlations.  The fact that one
of the three phases in the Blume-Capel model (the ``0'' phase) is
distinguishable from the other two at the tricritical point seems to
be crucial. From the joint distribution $p(M,Q)$, with $Q=\sum_is_i^2$
(figure~\ref{fig:contours}), one sees that there is no direct phase
space path between the ``plus'' and ``minus'' phases. In order to
migrate from one of these phases to the another, the system must pass
through the ``0'' phase.  It seems that the tricritical clusters of
``0''s are very sensitive to correlations in the PRNs, and that
``resonant enhancement'' of the triplet correlations occurs when the
system size is a multiple of the larger generator lag. The critical
clusters in the Potts model are much less susceptible to these
specific correlations.

\section{Conclusions}

To summarize, we have demonstrated that the interplay between the
physical properties of a system (principally the system size) and
random number generator triplet correlations can lead to strong
systematic errors in simulations even when simple Metropolis updating
is used. Such ``resonant enhancement'' of PRN correlations can be very
specific to a particular situation, as in our case of the tricritical
Blume-Capel model, and not be observable at all in other, albeit very
similar systems.  Our observation therefore supports the view that a
random number generator is only ``good'' in conjunction with the
specific application for which it has been tested -- every new
application is also a new pseudo random number test, where known and
unknown correlations may gain importance in an uncontrolled way.

{\bf Acknowledgments} We are grateful to D.P. Landau and K. Binder for
useful discussions, and to P. Nielaba for practical help with some of
the tests. Some of our simulations were carried out using externally
provided versions of the R250, e.g. on the CRAY-YMP at the
H\"ochstleistungs-rechenzentrum J\"ulich.

\begin{figure}[h]

\caption{(a) Magnetisation distribution of the Blume Capel model on the
$V=500$ site FCC lattice at $k_B T=3, \Delta=5.67$, obtained
using serial updating with the R250 generator. (b) The same
distribution obtained using random site updating.}
\label{fig:asymm}
\end{figure}

\end{document}